\documentstyle[prd,preprint,aps]{revtex}
\begin{document}
\draft
\title{Tau Polarizations in the Three-body Slepton Decays with
Stau as the NLSP.}
\author{Chih-Lung Chou\footnote{choucl@phys.sinica.edu.tw} and
Chung-Hsien Chou\footnote{chouch@phys.sinica.edu.tw}}
\address{Institute of Physics, Academia Sinica, Nankang, Taipei,
Taiwan 11529}
\maketitle
\begin{abstract}
In the gauge-mediated supersymmetry breaking models with scalar
tau as the next-to-lightest supersymmetric particle, a scalar
lepton may decay dominantly into its superpartner, tau lepton, and
the lightest scalar tau particle through $\tilde \ell \rightarrow
\ell \tau^{\pm} {\tilde \tau_1^{\mp}}$. We give detailed formulas
for the three-body decay amplitudes and the polarization asymmetry
of the outgoing tau lepton . We find that the tau polarizations
are sensitive to the model parameters such as the stau mixing
angle, the neutralino to slepton mass ratio and the neutralino
mixing
effect. 

\end{abstract}

\section{Introduction}
Supersymmetry provides a natural solution to the hierarchy problem
that associates with weak scale physics and Planck scale physics
\cite{coupling}. If nature is supersymmetric, it must
spontaneously break supersymmetry and transmit the message to low
energy effective models.  Two approaches are widely considered to
satisfy the message transmitting strategy. One is the gravity
mediation approach which has the supersymmetry breaking scale
$\sqrt{F}$ about of order $10^{11}$ GeV\cite{sugra,sugra02}.  The
gravitino mass is typically much larger than those of gauginos and
sfermions under this framework. Due to the largeness of
$\sqrt{F}$, the gravitino interaction is also too weak to play
important roles in collider phenomenology. Another approach uses
gauge interactions to mediate supersymmetry breaking
effect\cite{GMSB01,GMSB02}. In the gauge-mediated supersymmetry
breaking (GMSB) scenarios, the supersymmetry breaking scale
$\sqrt{F}$ is sufficiently small so that the gravitino $\tilde G$
is always the lightest supersymmetric particle (LSP) with mass

\begin{equation}
    m_{\tilde G}={F \over {\sqrt{3} M_{Pl}}}=2.37 \times 10^{-4} \times
    \left({\sqrt{F} \over 1 {\sf TeV}}\right)^2  \hspace{0.4cm} \sf{eV},
\label{eqn:gravitinomass}
\end{equation}

\noindent where $M_{Pl}$ denotes the reduced Planck scale $\sim
2.4 \times 10^{18}$ GeV. Since the soft masses of other
superparticles are associated with the weak scale in the GMSB
models, within typical range of $\sqrt{F}$, the gravitino is far
lighter than other superparticles and becomes the LSP.

It is well known that different scenarios for the next-to-lightest
supersymmetric particle (NLSP) in the GMSB models could be crucial
in discovering supersymmetry signals in colliders and may lead to
different phenomenologies.  In the neutralino NLSP scenarios, the
lightest neutralino $\tilde N_1$ decays into a photon and a
gravitino which carries missing energy and escapes the
detectors\cite{neutralino}. If the SUSY breaking scale $\sqrt{F}$
is less than a few thousand TeV, the neutralino decay $\tilde N_1
\rightarrow \gamma \tilde G$ would be prompt enough to occur
within the detector. One can therefore use inclusive $\gamma
\gamma + \not\!\!E_T + X$ signals in detecting supersymmetry. In
the slepton NLSP scenarios, the lightest slepton can promptly
decay into its partner and a gravitino with a sufficiently small
scale $\sqrt{F}$\cite{slepton}. Depending on the scale $\sqrt{F}$,
the slepton NLSP may live long enough to leave tracks and kinks in
detectors. In general, other NLSP scenarios are also plausible
depending upon the various constructions of supersymmetry breaking
models.

Scalar leptons can be easily found at future linear $e^+ e^-$
colliders if it is kinematically allowed\cite{LC}. By measuring
their masses and decay distributions one could determine some of
the MSSM parameters. The phenomenology of sleptons in future
linear colliders  then deserves detailed study.

In this paper we concentrate on the GMSB scenarios in which the
lighter scalar tau lepton (stau) $\tilde \tau_1$ is the NLSP. In
general, we can write down the stau mass eigenstates $\tilde
\tau_1$ and $\tilde \tau_2$ in terms of the stau mixing angle
$\theta_{\tilde \tau}$

\begin{eqnarray}
{\tilde \tau_1}&=& \ \ \ \cos\theta_{\tilde \tau} {\tilde \tau_L}+
\sin\theta_{\tilde \tau}{\tilde \tau_R}  \label{eqn:stau1mass}\\
{\tilde \tau_2}&=&-\sin\theta_{\tilde \tau} {\tilde \tau_L}+
\cos\theta_{\tilde \tau} {\tilde \tau_R},
\end{eqnarray}

\noindent where $\theta_{\tilde \tau}$ ranges from $0 \leq
\theta_{\tilde \tau} < \pi$. Since the mixing is proportional to
fermion Yukawa coupling, due to the small mixing efffects for
scalar electrons (selectrons) and for scalar muons (smuons) , the
lightest selectron $\tilde e_1$ and the lightest smuon $\tilde
\mu_1$ are nearly right-handed sleptons with almost equal masses.
We will use $\tilde e_R$ and $\tilde \mu_R$ to represent the
lighter selectron and smuon respectively in this paper.

 $\tilde \tau_1$ can
decay into a tau lepton and a gravitino $\tilde G$ in the
detectors if the SUSY scale $\sqrt{F}$ is not too large. If the
scale $\sqrt{F}$ is sufficiently large, the stau NLSPs may live
long enough to leave tracks or kinks in the collider detectors. In
many of the GMSB models, the lighter selectron and the lighter
smuon are merely right-handed sleptons with nearly degenerate
masses. Unlike $\tilde \tau_1$, they can decay not only through
the two-body processes $\tilde e_R \rightarrow e \tilde G$ and
$\tilde \mu_R \rightarrow \mu \tilde G$, but also through the
processes such as $\tilde \ell_R \rightarrow \ell \tilde N_1$,
$\tilde \ell_R \rightarrow \nu_{\ell} \bar {\nu_\tau} \tilde
\tau_1$, and $\tilde \ell_R \rightarrow \ell \tau \tilde \tau_1$
depending upon the model parameters. For instance, $\tilde \ell_R
\rightarrow \ell \tau \tilde \tau_1$ will be kinematically
forbidden if the mass difference $(m_{\tilde{\ell}_R} - m_{\tilde
\tau_1})$ is smaller than the tau mass $m_{\tau}$. This could
happen in the so-called slepton co-NLSP
scenarios\cite{SUSYsignal}.

However, in the stau NLSP scenarios, there exists some parameter
space where the three-body processes $\tilde \ell_R \rightarrow
\ell \tau \tilde \tau_1$ could dominate over the two-body
processes $\tilde \ell_R \rightarrow \ell \tilde G$ and leads to
different signatures in colliders \cite{3bodyslepton}.

If the decay $\tilde{\tau}_1 \rightarrow \tau\tilde{G}$ talkes
place outside the detector, then one can directly identify the
stau tracks or decay kinks. Thus in the $e^+ e^-$ colliders, the
pair-produced sleptons may decay in collider detectors through
$e^+ e^- \rightarrow \tilde \ell_R^+ \tilde \ell_R^- \rightarrow
\ell^+ \ell^- \tau \tau \tilde \tau_1 \tilde \tau_1$. In the $e^-
e^-$ colliders, right-handed sleptons are produced through the
t-channel processes and then decay as $e^- e^- \rightarrow \tilde
\ell_R^- \tilde \ell_R^- \rightarrow \ell^- \ell^- \tau \tau
\tilde \tau_1 \tilde \tau_1 $. It is also found that $\tilde e_R$
and $\tilde \mu_R$ could have macroscopic decay lengths that are
smaller than the dimensions of typical detectors if the mass
difference $|m_{\tilde \ell_R}-m_{\tau}-m_{\tilde \tau_1}|$ is
suitably small. If this is the case, the secondary vertices of the
three-body processes could be resolved and each $\tilde \tau_1$
particle can be identified as coming from one of the pair-produced
selectrons or smuons. In this case one could measure the
polarizations of the $\tau$ leptons. The $\tau$  polarization
provides another observable which could help to pin down the stau
mixing angle and/or the ratio of the neutrilino  and the slepton
masses.

However, there might be the possibility that the two-body and
three-body decay modes are of comparable amplitudes and can
therefore decay in the detector. In this case we might have the
decay process $e^- e^+ \rightarrow \tilde \ell_R^- \tilde \ell_R^+
\rightarrow \ell^- \ell^+ \tau^{\pm} \tilde \tau_1^{\mp}\not\!\!E
 \rightarrow \ell^- \ell^+ \tau^{\pm}
\tau^{\mp}\not\!\!E\not\!\!E $, i.e., two tau leptons with
opposite charges in the final state.  One of the tau lepton is
coming from the three-body decay of the slepton therefore its
kinetic energy is less then $|m_{\tilde \ell_R}-m_{\tau}-m_{\tilde
\tau_1}| $. The other tau lepton is from the two-body decay of the
scalar tau and hence carries larger momentum and is much harder
than the tau lepton originated from the three-body decay. In this
interesting case we can easily identify the softer tau lepton from
the three-body decay from $\tilde \ell_R$.

Motivated by the above observation, we calculate the polarizations
for the tau leptons in the three-body decays $\tilde \ell_R
\rightarrow \ell \tau^{\mp} \tilde \tau_1^{\pm}$ in the paper. It
is expected that the tau polarizations should depend on the stau
mixing angle and the mass ratio $m_{\tilde N_1} / m_{\tilde e_R}$.
In the section two of the paper, we set the notations and give
formulas for the polarizations of the tau leptons in the
three-body slepton decays. Numerical results are presented and
discussed in the section three.  In the section four, we make our
conclusion.

\section{Three-Body slepton decays}
As mentioned in the previous section, the three-body decays
 $\tilde \ell_R \rightarrow \ell \tau \tilde \tau_1$ could dominate
 over the two-body decays $\tilde \ell_R \rightarrow \ell \tilde G$
 in some parameter space if the SUSY breaking scale $\sqrt{F}$ is
 sufficiently large. We restrict
 on the scenario where the two-body processes
 $\tilde e_R \rightarrow e+\tilde N_i$ are forbidden, i.e., we
 assume the following mass relation:

\begin{equation}
m_e+m_{\tilde N_i} > m_{\tilde e_R} > m_e+m_{\tau}+m_{\tilde
\tau_1}. \label{eqn:massrelation}
\end{equation}

\noindent Therefore the only competing decay processes for $\tilde
e_R$ and $\tilde \mu_R$ are $\tilde \ell_R \rightarrow \ell
\tau^{\pm} {\tilde \tau_1^{\mp}}$.

Because of the invariance of charge conjugation and parity
transformation, we can write down the following identities for the
selectron decays:

\begin{eqnarray}
\Gamma(\tilde e_R^+ \rightarrow e^+ \tau_R^{\pm} \tilde
\tau_1^{\mp}) = \Gamma(\tilde e_R^- \rightarrow e^- \tau_L^{\mp}
\tilde \tau_1^{\pm}) \label{eqn:CPsym01}  \\ \Gamma(\tilde e_R^+
\rightarrow e^+ \tau_L^{\pm} \tilde \tau_1^{\mp}) = \Gamma(\tilde
e_R^- \rightarrow e^- \tau_R^{\mp} \tilde \tau_1^{\pm}),
 \label{eqn:CPsym02}
\end{eqnarray}

\noindent where $L$ ($R$) denotes the left (right) helicity for
the tau leptons in the decay processes. Therefore we will only
discuss the decays of $\tilde e^+_R$.

At tree level, the selectron $\tilde e_R$ decays through the
propagation of bino $\tilde B$ and wino $\tilde W^3$. Since bino
$\tilde B$, wino $\tilde W^3$, and the two neutral higgsinos
$\tilde h^0_1$, $\tilde h^0_2$ can mix and form the neutralino
mass eigenstates $\tilde N_i$, it is appropriate to do calculation
in terms of the neutralino parameters. In the basis of $\psi_i=
(\tilde B, \tilde W^3, i \tilde h^0_1, i \tilde h^0_2)$, the
neutralino states and the associated diagonal mass matrix $M_D$
are given by

\begin{eqnarray}
\tilde N_i &=& \left( V^+ \right)_{ij} \psi_j \\ M_D &=& V^+ M  V.
\end{eqnarray}

\noindent Here $V$ denotes the transforming matrix which
diagonalizes the neutralino mass matrix $M$

\begin{eqnarray}
M&=&\left(
    \begin{array}{cccc}
    m_1 & 0 & m_{Z}\sin{\theta_W}\cos\beta & -m_Z \sin{\theta_W}\sin\beta \\
    0 & m_2 & -m_Z \cos{\theta_W}\cos\beta & m_Z \cos{\theta_W}\sin\beta \\
    m_{Z}\sin{\theta_W}\cos\beta & -m_Z \cos{\theta_W}\cos\beta & 0& -\mu \\
    -m_Z \sin{\theta_W}\sin\beta & m_Z \cos{\theta_W}\sin\beta & -\mu & 0
    \end{array}
    \right), \label{eqn:neumass}
\end{eqnarray}

\noindent where $m_Z$ is the mass of $Z$ boson, $m_1$ denotes the
soft breaking mass of bino $\tilde B$, $m_2$ is the soft breaking
mass of wino $\tilde W^3$, $\theta_W$ denotes the weak Weinberg
angle and $\tan\beta$ is the ratio between the vacuum expectation
values of the two Higgs doublets in the model, and $\mu$ is the
supersymmetric Higgs mass parameter.

In our study, the tau polarizations are measured in the lab frame.
When the collider beam energy is threshold to produce the
slectrons in pairs, the selectrons may travel with velocity $v$
before decay and thus experience a boost which is characterized by
the boost factor $\gamma = (v/c)/\sqrt{1-(v/c)^2}$. Fig.\
\ref{fig:boost} shows the selectron boost direction in the lab
frame. The moving direction of the outgoing tau lepton is assigned
the Z-direction and the XZ plane is spanned by the outgoing
electron and the tau lepton. By summing all tree-level
contributions for the decay $\tilde e^+_R \rightarrow e^+ \tau^-
\tilde \tau_1^+$, the transition amplitudes $|i\mbox{\Large
m}^-_{(L,R)}|^2$ for different helicities $(L, R)$ of the tau
final states are given by:

\begin{eqnarray}
|i\mbox{\Large m}^-_{(R)}|^2 &=& |\sum_{j=1}^4 {2Q_e
V^*_{Rj}\sqrt{E_e} \over (q^2-m^2_{\tilde N_j})}
\{(C_{1j}E_{\tilde e_R}\sqrt{E+K} - C^*_{2j} m_{\tilde N_j}
\sqrt{E-K}) cos{\theta \over 2} \nonumber
\\ &+&C_{1j}P_{\tilde e_R}\sqrt{E+K}(\sin\phi \sin\varphi \sin{\theta
\over 2}+\sin\phi \cos\varphi \cos{\theta \over 2} -i \cos\phi
\sin{\theta \over 2})\} |^2 \label{eqn:AmpRN} \\ |i\mbox{\Large
m}^-_{(L)}|^2 &=& |\sum_{j=1}^4 {2Q_eV^*_{Rj}\sqrt{E_e} \over
 (q^2-m^2_{\tilde N_j})} \{(C_{1j}E_{\tilde e_R}\sqrt{E-K} -
 C^*_{2j} m_{\tilde N_j} \sqrt{E+K}) \sin{\theta \over2} \nonumber
 \\ &+&C_{1j}P_{\tilde e_R}\sqrt{E-K}(\sin\phi\sin\varphi \cos{\theta
 \over 2} - \sin\phi \cos\varphi \sin{\theta \over 2}+ i\cos\phi
 \cos{\theta \over 2})\} |^2
 \label{eqn:AmpLN}
\end{eqnarray}

\noindent where
\begin{equation}
q^2\equiv m^2_{\tilde e_R}-2PE_{\tilde e_R}-2PP_{\tilde
e_R}\sin\phi\cos(\theta - \varphi), \label{eqn:qMomentum}
\end{equation}

\noindent $(E_{\tilde e_R}, P_{\tilde e_R})$ and $(E, K)$ denote
the (energy, momentum) pairs of the right-handed selectron and the
outgoing tau lepton in the lab frame respectively, $P$ is the
outgoing electron energy, $m_{\tilde e_R}$ and $m_{\tilde N_j}$
denote the masses of the right-handed selectron and the $j$th
neutralino respectively, $Q_e$ represents the electric coupling of
electron, and the coefficients $C_{1j}, C_{2j}$, and $V_{Rj}$ are
given by\cite{sleptonProd}

\begin{eqnarray}
V_{Lj}&=&{1\over 2\sin\theta_W}V_{2j}+{1\over 2\cos\theta_W}V_{1j}
\label{eqn:VL} \\ V_{Rj}&=&{1\over \cos\theta_W}V_{1j}
\label{eqn:VR}\\ C_{1j}&=& \lambda_{\tau} V_{3j}
\cos\theta_{\tilde \tau} -{\sqrt 2}Q_e V_{Rj} \sin\theta_{\tilde
\tau} \label{eqn:C1}\\ C_{2j}&=& \lambda_{\tau} V_{3j}
\sin\theta_{\tilde \tau} +{\sqrt 2}Q_e V_{Lj} \cos\theta_{\tilde
\tau} \label{eqn:C2}.
\end{eqnarray}

\noindent Here $\lambda_{\tau}$ is the coupling strength for the
 tau Yukawa coupling term $\lambda_{\tau} H_1 L_{\tau} R_{\tau}$.
Similarly, the decay process $\tilde e^+_R \rightarrow e^+ \tau^+
\tilde \tau_1^-$ has the transition amplitudes

\begin{eqnarray}
|i\mbox{\Large m}^+_{(R)}|^2&=&|\sum_{j=1}^4 {2Q_e
V^*_{Rj}\sqrt{E_e} \over (q^2-m^2_{\tilde N_j})}
\{(-C^*_{1j}m_{\tilde N_j}\sqrt{E-K} + C_{2j} E_{\tilde e_R}
\sqrt{E+K}) \cos{\theta \over2} \nonumber
\\ &+&C_{2j}P_{\tilde e_R}\sqrt{E+K}(\sin\phi \sin\varphi \sin{\theta
\over 2}+\sin\phi \cos\varphi \cos{\theta \over 2} -i \cos\phi
\sin{\theta \over 2})\}|^2
 \label{eqn:AmpRP} \\
 |i\mbox{\Large m}^+_{(L)}|^2&=&|\sum_{j=1}^4 {2Q_e V^*_{Rj}\sqrt{E_e} \over
(q^2-m^2_{\tilde N_j})} \{ (-C^*_{1j}m_{\tilde N_j}\sqrt{E+K} +
C_{2j} E_{\tilde e_R} \sqrt{E-K}) \sin{\theta \over2} \nonumber
\\&+&C_{2j} P_{\tilde e_R}\sqrt{E-K}(\sin\phi\sin\varphi
\cos{\theta \over 2}
 - \sin\phi \cos\varphi \sin{\theta \over 2}+ i\cos\phi \cos{\theta
 \over 2})\} |^2. \label{eqn:AmpLP}
\end{eqnarray}

\noindent As easily seen from Eqs.\
(\ref{eqn:AmpRN}-\ref{eqn:AmpLP}), the amplitudes do depend upon
the selectron boost angles and the selectron momentum $P_{\tilde
e_R}$.

Taking into account the selectron boost effect, the decays rate
$\Gamma_{L, R}$ of different tau lepton helicities L or R for
$\tilde e_R$ is obtained as

\begin{equation}
\Gamma_{(L, R)}=\int d\phi d\varphi F(\phi,
\varphi)\Gamma_{(L,R)}(\phi,\varphi) , \label{eqn:decayrate}
\end{equation}

\noindent with
\begin{eqnarray}
\Gamma_{(L,R)}(\phi,\varphi)&\equiv&\int {dE d[\cos\theta]\over
{64 \pi^3 E_{\tilde e_R}}}{KP|i\mbox{\Large m}_{(L,R)}|^2 \over
{{E_{\tilde e_R}-E+K \cos\theta - P_{\tilde e_R}
\sin\phi(\sin\varphi\sin\theta+\cos\varphi\cos\theta)}}}
\label{eqn:rateNangle} \\
 P&\equiv& {{m^2_{\tilde e_R}+m^2_{\tau}-m^2_{\tilde \tau_1}-2E
 E_{\tilde e_R}+2P_{\tilde e_R}K\sin\phi\cos\varphi} \over
 {2[E_{\tilde e_R}-E+Kcos\theta - P_{\tilde
e_R}\sin\phi(\sin\varphi\sin\theta+\cos\varphi\cos\theta)
]}},\label{eqn:Pformula}
\end{eqnarray}

\noindent where $F(\phi,\varphi)$ is the probability density
function for selectron boosting in the $(\phi, \varphi)$
direction, $E$ is the energy for the outcoming tau lepton,
$K=\sqrt{(E^2-m^2_{\tau})}$ denotes the momentum of the tau
lepton, $P$ is the energy of the outcoming electron, and $\theta$
is the angle between the tau and the electron.

Naively, the decay rates with selectron boost effects may differ
to those without selectron boost effects to the order of
$O(\gamma)$, i.e. $O(P_{\tilde e_R} / E_{\tilde e_R})$. On the
other hand, the spinless nature of selectrons leads to the
following symmetry for $F(\phi, \varphi)$:

\begin{equation}
F(\phi,\pm\varphi)=F(\phi,\pi \pm \varphi)=F(\pi-\phi,
\pm\varphi)=F(\pi-\phi, \pi\pm\varphi). \label{eqn:Fsym}
\end{equation}

\noindent Thus all terms that are linear in $\cos\varphi$ or
$\sin\varphi$ should be cancelled after integrating out the
$\varphi$ dependence. Consequently, by observing that

\begin{equation}
\sqrt{E_{rest} \pm K_{rest}} = \sqrt{E \pm K} \{1 \mp {1\over
2}\gamma \cos\varphi \sin\phi + O(\gamma^2) \}
\end{equation}

\noindent where $E_{rest}$ ($K_{rest}$) denotes the energy
(momentum) of the outgoing tau lepton in the selectron rest frame,
the decay rates in Eq.(\ref{eqn:decayrate}) are actually differ to
those without selectron boost effects to the order of
$O(\gamma^2)$. Therefore, we simply do all the calculations in the
selectron rest frame in the rest of the paper.

\section{Tau Lepton Polarization}
The tau lepton polarization asymmetry can be defined as

\begin{equation}
P_{\tau} \equiv {<\Gamma_R>-<\Gamma_L> \over
<\Gamma_R>+<\Gamma_L>}.
\end{equation}

\noindent We can easily calculate the asymmetry by using the
formulas presented in the previous section. From Eq.s\
(\ref{eqn:CPsym01}) and (\ref{eqn:CPsym02}) the CP invariance
leads to the following relations for the tau polarization:

\begin{eqnarray}
P_{\tau}(\tilde e_R^+ \rightarrow e^+ \tau^+ \tilde \tau_1^-) =
-P_{\tau}(\tilde e_R^- \rightarrow e^- \tau^- \tilde \tau_1^+)
\nonumber \\
P_{\tau}(\tilde e_R^+ \rightarrow e^+ \tau^- \tilde
\tau_1^+) = -P_{\tau}(\tilde e_R^- \rightarrow e^- \tau^+ \tilde
\tau_1^-) \label{Polarelation}
\end{eqnarray}

\noindent Therefore we calculate only $P_{\tau}(\tilde e_R^+
\rightarrow e^+ \tau^+ \tilde \tau_1^-)$ and $P_{\tau}(\tilde
e_R^+ \rightarrow e^+ \tau^- \tilde \tau_1^+)$ and give the
numerical results in this section.

Generally, the transistion amplitudes and the associated decay
rates presented in the previous section are sensitive to the
neutralino masses and mixing effects. For instance, by assuming
the dominance of $\tilde N_1$ propagating in the tree-level decay
processes in the Bino-like lightest neutralino scenario, the
amplitude functions in the limit $\Delta \equiv (m_{\tilde
e_R}-m_{\tilde \tau_1}) << m_{\tilde e_R}$ are obtained
approximately as

\begin{eqnarray}
|i\mbox{\Large m}^-_{(R)}|^2&\propto& \{E_e |m_{\tilde e_R}
\sqrt{E+K} \sin\theta_{\tilde \tau}+ {m_{\tilde B} \over
2}\sqrt{E-K} \cos\theta_{\tilde \tau}|^2 \cos^2{\theta \over
2}\},\label{R-formula}\\ |i\mbox{\Large m}^+_{(R)}|^2&\propto&
\{E_e |m_{\tilde B}\sqrt{E-K} \sin\theta_{\tilde\tau}+ {m_{\tilde
e_R} \over 2}\sqrt{E+K} \cos\theta_{\tilde \tau}|^2 \cos^2{\theta
\over 2}\}, \label{R+formula}\\ |i\mbox{\Large m}^-_{(L)}|^2
&\propto& \{E_e |m_{\tilde e_R}\sqrt{E-K}\sin\theta_{\tilde\tau} +
{m_{\tilde B}\over 2}\sqrt{E+K}\cos\theta_{\tilde \tau}|^2
\sin^2{\theta \over 2}\}, \label{L-formula}
\\ |i\mbox{\Large m}^+_{(L)}|^2 &\propto& \{E_e |m_{\tilde B}
\sqrt{E+K}\sin\theta_{\tilde
\tau}+{m_{\tilde e_R}\over 2}\sqrt{E-K}\cos\theta_{\tilde \tau}|^2
\sin^2{\theta \over 2} \} , \label{L+formula}.
\end{eqnarray}

\noindent From Eq.s\ (\ref{R-formula}-\ref{L+formula}), it is
easily seen that the polarization ratio $P_{\tau}$ is determined
by the mass ratio $m_{\tilde B}/m_{\tilde e_R}$ . In the purely
right-handed $\tilde \tau_1$ limit $(\sin\theta_{\tilde \tau}=1)$,
$P_{\tau}$ becomes

\begin{equation}
P_{\tau}\approx \pm {\int^{\Delta}_{m_{\tau}} dE
\hspace{0.1cm}(E^2-m_{\tau}^2)(\Delta-E)^2 \over
\int^{\Delta}_{m_{\tau}} dE \hspace{0.1cm}E
\sqrt{E^2-m_{\tau}^2}(\Delta-E)^2}, \hspace{2cm}(\pm)
\hspace{0.2cm} {\sf for} \hspace{0.2cm} \tau^{\mp}
\hspace{0.2cm}{\sf final}\hspace{0.2cm} {\sf states}.
\label{eqn:approasym}
\end{equation}

\noindent Eq.\ (\ref{eqn:approasym}) shows that the $P_{\tau}$
ratios always have opposite signs for $\tau^+$ and $\tau^-$
production in the selectron decays, i.e., the $\tilde e^+_R
\rightarrow e^+ \tau^- \tilde \tau^+_1$ process tends to have
right-handed tau leptons in the final states and the $\tilde e^+_R
\rightarrow e^+ \tau^+ \tilde \tau^-_1$ process tends to produce
left-handed $\tau^+$'s in this limit.

The tau-lepton polarizations as the functions of the mass
difference $\Delta $ with different $\cos\theta_{\tilde \tau}$
values are illustrated in Fig.\ \ref{fig:P2Delta} (a) and (b)
where the  mass ratio $r_{\tilde B}\equiv m_{\tilde B}/m_{\tilde
e_R}=2$  is kept fixed. Although it was pointed out in
\cite{3bodyslepton} that when the mass differences between $\tilde
\tau_1$ and $\tilde e_R$ and $\tilde \mu_R$ are less than 10 GeV,
the range of $\cos{\theta_{\tilde{\tau}}}$ will be within $0.1\leq
|\cos{\theta_{\tilde \tau}}| \leq 0.3$ , we find that if we assume
that $m_{\tilde e_R} \sim 200$ GeV the mass differences between
$\tilde \tau_1$ and $\tilde e_R$ and $\tilde \mu_R$ can be up to
20 GeV while keeping $0.1\leq |\cos{\theta_{\tilde \tau}}| \leq
0.3$. As shown in the plot (a), $P(\tau^+) \equiv P_{\tau}(\tilde
e_R^+ \rightarrow e^+ \tau^+ \tilde \tau_1^-)$ decreases as
$\Delta$ increases. For a given $\Delta$ value, $P( \tau^+)$
always has larger values for larger $\cos\theta_{\tilde \tau}$
values. This could be explained by the fact that when
$\cos\theta_{\tilde \tau}$ increases, the amplitude
$|i\mbox{\Large m}^+_{(R)}|^2$ always increases faster than
$|i\mbox{\Large m}^+_{(L)}|^2$ does when the mixing angle are
within the range $0.1\leq |\cos{\theta_{\tilde \tau}}| \leq 0.3$ .
Although not shown in the figure, from Eq.s\
(\ref{R-formula},\ref{L-formula}) one can easily see that when
$\Delta$ is kept fixed, a larger $\cos{\theta_{\tilde \tau}}$ will
give smaller polarization $P(\tau^-)\equiv P_{\tau}(\tilde e_R^+
\rightarrow e^+ \tau^- \tilde \tau_1^+)$.

 As shown in Fig.\ \ref{fig:P2Delta}(b), the ratio between
  $P(\tau^+)$ and $P(\tau^-)$ is a slow varying function
  of $\Delta$ with fixed $r_{\tilde B}=2$ for most of the
$\cos\theta_{\tilde \tau}$ range except for $\cos\theta_{\tilde
\tau}=-0.3$. This is due to the complex interplay between the
factors $\sqrt{E+K}\cos\theta_{\tilde\tau}$ and
$\sqrt{E-K}\sin\theta_{\tilde\tau}$.  When the mass difference
$\Delta$ is small, $\sqrt{E+K}$ and $\sqrt{E-K}$ are comparable
and both $|i\mbox{\Large m}^+_{(R)}|^2$ and $|i\mbox{\Large
m}^-_{(L)}|^2$ are getting smaller as $\cos\theta_{\tilde \tau}$
changing from $-0.1$ to $-0.3$. On the other hand, when $\Delta$
goes larger, say $\Delta=20$ GeV, $\sqrt{E+K}$ is much larger than
$\sqrt{E-K}$ and $|i\mbox{\Large m}^-_{(L)}|^2$ ($|i\mbox{\Large
m}^+_{(R)}|^2$) is getting larger (smaller) as $\cos\theta_{\tilde
\tau}$ changing from $-0.1$ to $-0.3$. It thus leads to the
intersections for those curves with  negative $\cos\theta_{\tilde
\tau}$ values.

Fig.s\ \ref{fig:P2Ratio}(a) and \ref{fig:P2Ratio}(b) show the mass
ratio $r_{\tilde B}$ dependence of the tau polarizations. As shown
in Fig.\ \ref{fig:P2Ratio}(a), $P(\tau^+)$ is a slow varying
function of $r_{\tilde B}$ with fixed $\Delta$ value ($\Delta =
\mbox { 20 GeV}$ in the plot). However, the ratio $P(\tau^+
)/P(\tau^-)$ varies significantly for $0.1 \lesssim
\cos\theta_{\tilde \tau} \lesssim 0.3$ with fixed $\Delta$ value
as shown in \ref{fig:P2Ratio}(b). This plot may help in
determining the model parameters such as $r_{\tilde B}$ and
$\cos\theta_{\tilde \tau}$ in the Bino-like lightest neutralino
scenarios. For example, a ratio $ P(\tau^+ )/P(\tau^-)= -1.5$ will
lead to $\cos\theta_{\tilde \tau}\simeq 0.3$ and $r_{\tilde
B}\gtrsim 2.5$ in the model with $\Delta = 20\mbox{ GeV}$.

In Fig.s\ \ref{fig:P2Mixing}(a-c) we plot the tau polarization
functions and their ratio as functions of $\cos\theta_{\tilde
\tau}$ for $r_{\tilde B}=2$ and five choices $\Delta = 4, 8, 12,
16$, and $20 \mbox{ GeV}$. Fig.\ \ref{fig:P2Mixing}(a) shows that
the polarization $P(\tau^+)$ increases as $\cos\theta_{\tilde
\tau}$ increases for the $\cos\theta_{\tilde \tau}$ range under
consideration. Similar behaviors occur when varying the $\Delta$
value in the model. Usually a larger $\Delta$ value will imply a
lower tau polarization for $(\tilde e_R^+ \rightarrow e^+ \tau^+
\tilde \tau_1^-)$ and a higher tau polarization for $(\tilde e_R^+
\rightarrow e^+ \tau^- \tilde \tau_1^+)$. Several curves for the
tau polarization ratio $P(\tau^+)/P(\tau^-)$ are plotted against
the stau mixing parameter $\cos\theta_{\tilde \tau}$ with
different $\Delta$ values in Fig.\ \ref{fig:P2Mixing}(b). Varying
the $r_{\tilde B}$ parameter could dramatically change the
$\cos\theta_{\tilde \tau}$ dependence of $P(\tau^+)/P(\tau^-)$. We
plot $P(\tau^+)/P(\tau^-)$ for five choices $r_{\tilde B}=1.0,
1.5, 2.0, 2.5$ and $3.0$ in Fig.\ \ref{fig:P2Mixing}(c). As shown
in the plot, the polarization ratio $P(\tau^+)/P(\tau^-)$ depends
on $r_{\tilde{B}}$ sensitively when $0.3 \gtrsim
|\cos\theta_{\tilde \tau}| \gtrsim 0.2$.

We can also study the neutralino mixing effects on the tau-lepton
polarizations. Fig.\ \ref{fig:Higgsino} shows typical plots in
both the bino-like lightest neutralino scenario and in the
Higgsino-mixed neutralino region with the model parameters
$m_{\tilde e_R}=200$, $|\mu|=300$, $m_1=400$, $m_2=800 \mbox{
GeV}$ and $\tan\beta=10$. By comparing Fig.\ \ref{fig:Higgsino} to
\ref{fig:P2Ratio}(a), we find that the tau polarization behavior
in the Higgsino-mixed  neutralino scenarios may look similar to
that in the bino-like lightest neutralino scenarios with a larger
$|\cos\theta_{\tilde \tau}|$ value. For instance, the tau
polarization in the Higgsino-mixed neutralino region with
$\cos\theta_{\tilde\tau}=-0.1$ would look like the tau
polarization in the bino-dominant lightest neutralino scenario
with $\cos\theta_{\tilde\tau}\approx -0.3$.

In Fig.\ \ref{fig:pT} we plot the normalized distributions for the
final state electrons and tau leptons as functions of their
transverse momenta $p_T$ in the selectron rest frame. Here we
choose the model parameters $m_{\tilde{e}_R}=200$ GeV ,
$r_{\tilde{B}}=1.2$, $\cos{\theta_{\tilde{\tau}}}=-0.15$. We plot
three different values of $\Delta=12, 16, 20$ GeV respectively. As
seen from the figure, the peak values occur at about $p_T = 7.5,
5.8, 4.3$ GeV ( 6.0, 4.9, 3.7 GeV) for the final state tau lepton
(electron) distributions as we vary the $\Delta$ from 20, 16, to
12 GeV accordingly. It should be noticed that when the mass
difference $\Delta$ is too small, the final state tau may be too
soft to detect and the determination of the tau polarization would
be a challenge. However, if the mass splitting is large enough,
the polarization of the final state tau lepton can be measured.

According to the numerical results the knowledge of the tau
polarization in the three-body slepton decays can provide another
observable and may help to reveal the stau mixing angle and/or the
bino-slepton mass ratio $m_{\tilde B}/m_{\tilde e_R}$ in the small
mass difference $\Delta << m_{\tilde e_R}$ limit. Experimentally,
there may be a way for measuring the polarized tau-lepton
productions in the selectron decay by observing its subsequent tau
decays $\tau^- \rightarrow \pi^- \nu_{\tau}$ and $\tau^+
\rightarrow \pi^+ \bar \nu_{\tau}$. The tau polarizations can be
obtained by measuring the spectrum of
$E_{\pi^{\pm}}/E_{\tau^{\pm}}$ in the decays.

\section{Conclusion}
In this paper, we investigated the tau polarizations for the
three-body slepton decays under the stau NLSP scenarios in the
small mass difference limit $\Delta << m_{\tilde e_R}$. The exact
formulas for calculating the slepton decay rates at the tree level
are provided in section two.

The measurements of tau lepton polarizations in the slepton decays
provide another window for studying the SUSY model parameters. In
the right-handed $\tilde \tau_R$ limit, the tau asymmetry ratio
$P_{\tau}$ is solely determined by the mass difference parameter
$\Delta$.  Numerical studies show explicit dependence on the stau
mixing angles and/or the mass ratio $m_{\tilde B}/m_{\tilde e_R}$.

We also studied the distributions of the final state electrons and
tau leptons as functions of their transverse momenta $p_T$ in the
selectron rest frame. The energy spectrum of the final state
leptons are mainly determined by the mass difference ratio $\Delta
\equiv m_{\tilde{e}_R}-m_{\tilde{\tau}_1}$. In our chosen
parameters $m_{\tilde{e}_R}=200$ GeV, $r_{\tilde{B}}=1.2$,
$\cos{\theta_{\tilde{\tau}}}=-0.15$ as we vary $\Delta$ from 12,
16 to 20 GeV, the peaks of the distributions range from $p_T=$
4.3, 5.8 to 7.5 GeV respectively. However if the mass difference
is too small, the final state tau lepton will to too soft and the
measurement of the polarization would be very difficult.  But if
$\Delta$ is large enough such that we can measure its
polarization, the knowledge of tau-lepton polarization may help in
determining the parameter space of a typical GMSB model.

\subsection*{Acknowledgments}

C.-L. Chou thanks M. E. Peskin for introducing the topic and
useful discussion in the early stage of this work. C.-H. Chou
thanks S.-C. Lee for helpful discussions. This work was supported
in part by the National Science Council of Taiwan(Grant No.
NSC89-2112-M-001-070 and NSC89-2811-M-001-0100).


\begin{figure}
\caption{The selectron boost direction $(\phi, \varphi)$ in the
lab frame where the electron moving direction and the tau moving
direction spans the XZ plane. Here $\theta$ denotes the angle
between the outgoing electron and the outgoing tau lepton.}
\label{fig:boost}
\end{figure}

\begin{figure}
\caption{Tau polarization as the function of $\Delta$ in the
three-body decays of $\tilde e_R$ with $r_{\tilde B}=2$ and
$\cos\theta_{\tilde \tau}$ = -0.3, -0.2, -0.1, 0.1, 0.2, 0.3. (a)
$P_{\tau}(e^+_{\tilde R} \rightarrow e^+ \tau^+ \tilde \tau_1^-)$
as function of $\Delta$. (b) The $P(\tau^+)/P(\tau^-)$ ratio as
 function of $\Delta$.} \label{fig:P2Delta}
\end{figure}

\begin{figure}
\caption{Tau polarization as the function of the mass ratio
$r_{\tilde B}\equiv m_{\tilde B}/m_{\tilde e_R}$ with seven
choices $\cos\theta_{\tilde \tau}$ = -0.3, -0.2, -0.1, 0.1, 0.2,
0.3 and  $\Delta=20$ GeV. (a) $P_{\tau}(e^+_{\tilde R} \rightarrow
e^+ \tau^+ \tilde \tau_1^-)$ as function of $r_{\tilde B}$. (b)
The $P(\tau^+)/P(\tau^-)$ ratio as function of $r_{\tilde B}$.}
\label{fig:P2Ratio}
\end{figure}

\begin{figure}
\caption{(a) Tau polarization as function of $\cos\theta_{\tilde
\tau}$ for $r_{\tilde B}=2$ and five choices $\Delta$ = 4, 8, 12,
16 and 20 GeV for the selectron decay $e^+_{\tilde R} \rightarrow
e^+ \tau^+ \tilde \tau_1^-$. (b) The ratio $P(\tau^+)/P(\tau^-)$
as function of $\cos\theta_{\tilde \tau}$ for $r_{\tilde B}=2$ and
five choices $\Delta$ = 4, 8, 12, 16 and 20 GeV. (c)The ratio
$P(\tau^+)/P(\tau^-)$ for $\Delta$ = 20 GeV and five choices
$r_{\tilde B}$ = 1.0, 1.5, 2.0, 2.5, 3.0.} \label{fig:P2Mixing}
\end{figure}

\begin{figure}
\caption{The dependence of tau polarization on the mass difference
 $\Delta \equiv m_{\tilde e_R} - m_{\tilde \tau_1}$ with different $\cos\theta_{\tilde
\tau}$ values. Curves are plotted both in the bino-like neutralino
scenario and in the higgisino-like neutralino scenario for the
decay process $e^+_{\tilde R} \rightarrow e^+ \tau^+ \tilde
\tau_1^-$. Here we choose the model parameters $m_{\tilde
e_R}=200$, $|\mu|=300$, $m_1=400$, $m_2=800 \mbox{ GeV}$ and
$\tan\beta=10$. }\label{fig:Higgsino}
\end{figure}

\begin{figure}
\caption{Lepton $p_T$ distributions in the rest frame of the
$\tilde{e}_R^+$ decaying to $e^+\tau^+\tilde{\tau}^-$. Normalized
$p_T$ distributions for final $e$ (dot-dash lines) and $\tau$
(solid lines) for the model parameters with $m_{\tilde{e}_R}=200$
GeV, $\cos{\theta_{\tilde{\tau}}}=-0.15$, $r_{\tilde{B}}=1.2$, and
$\Delta=12, 16, 20$ GeV respectively. }\label{fig:pT}
\end{figure}

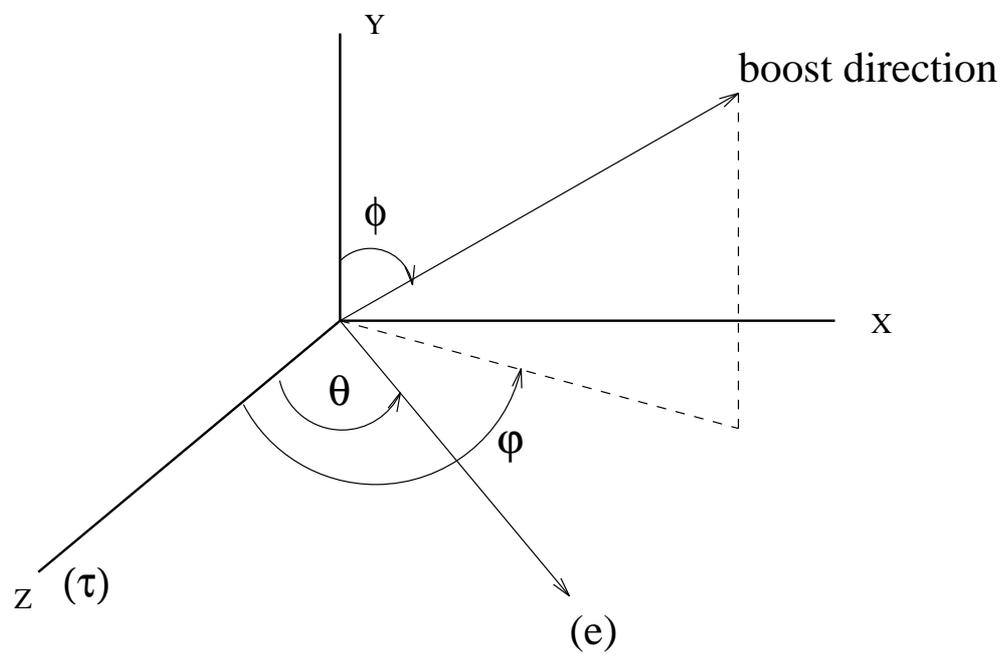
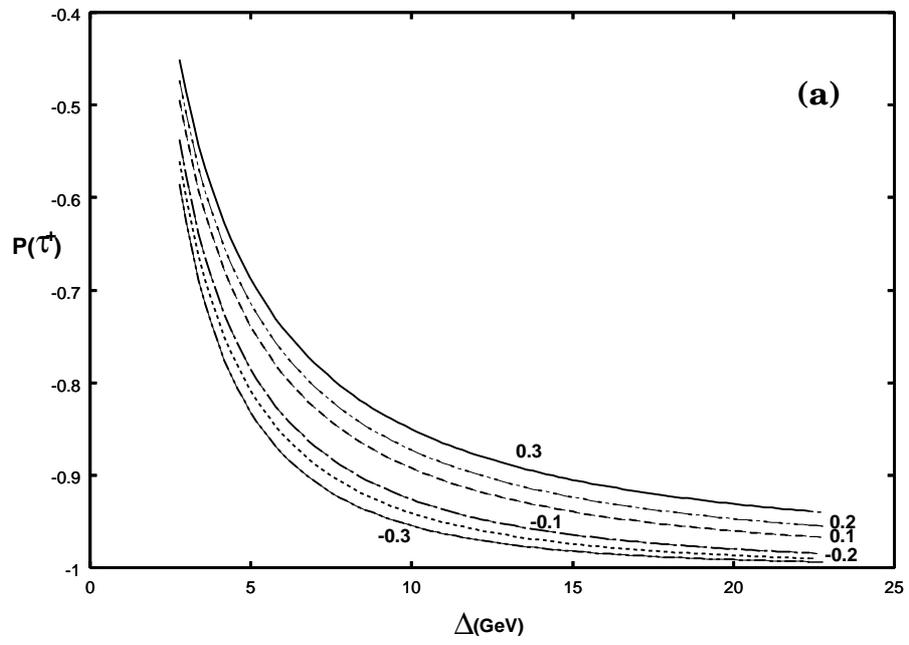
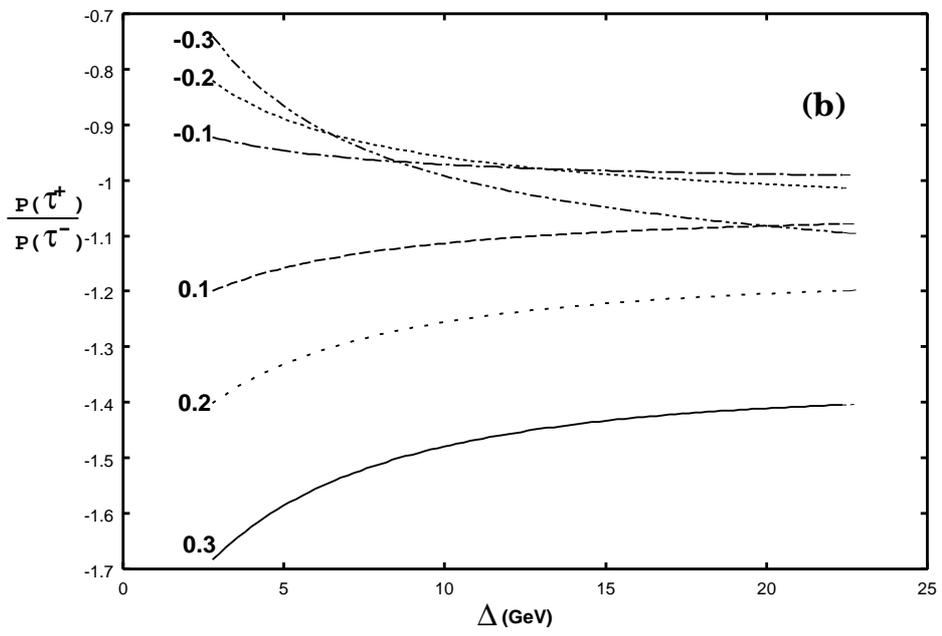
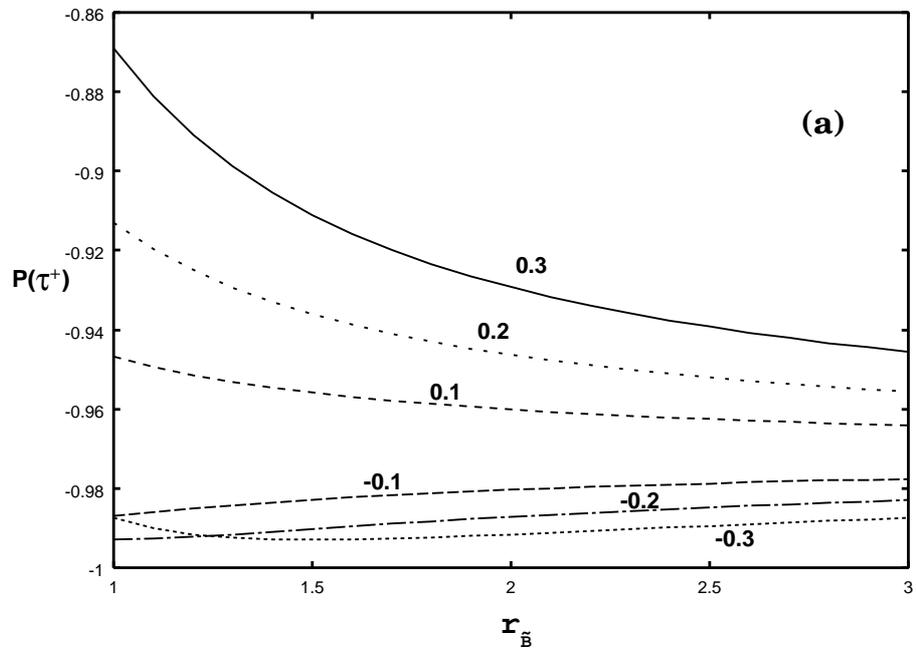
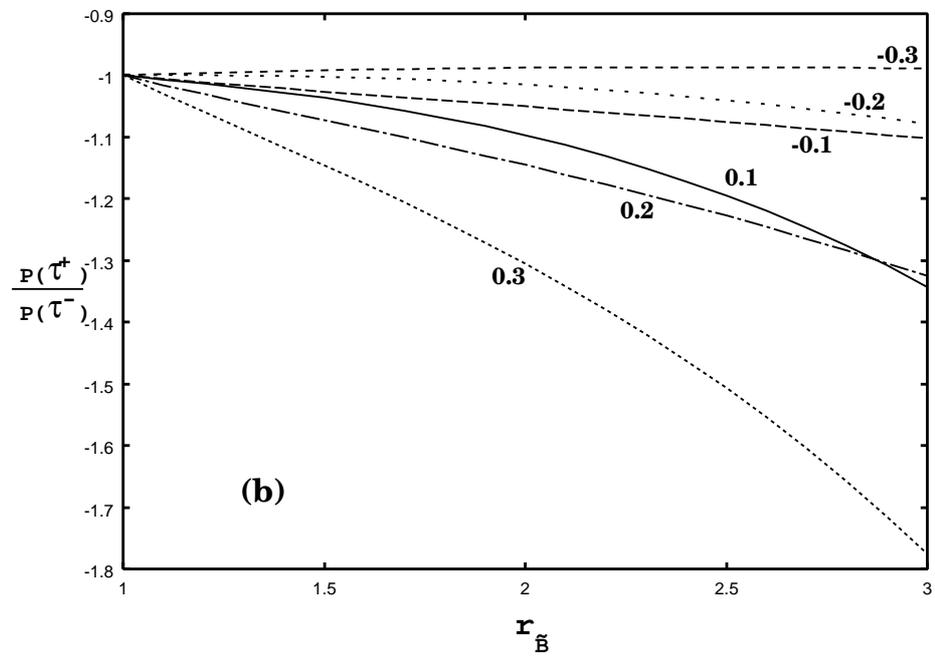
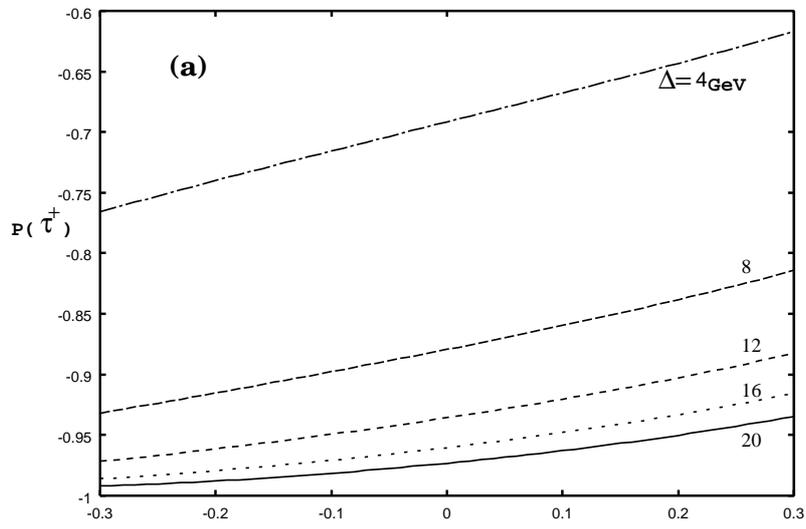
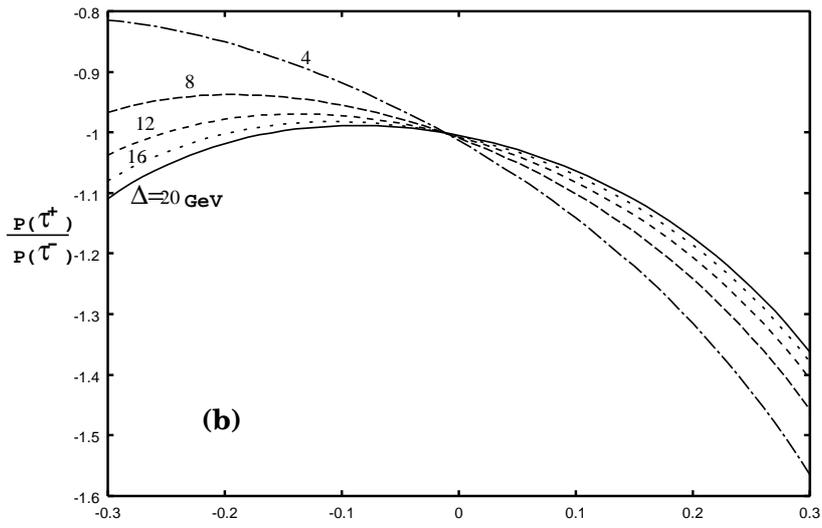
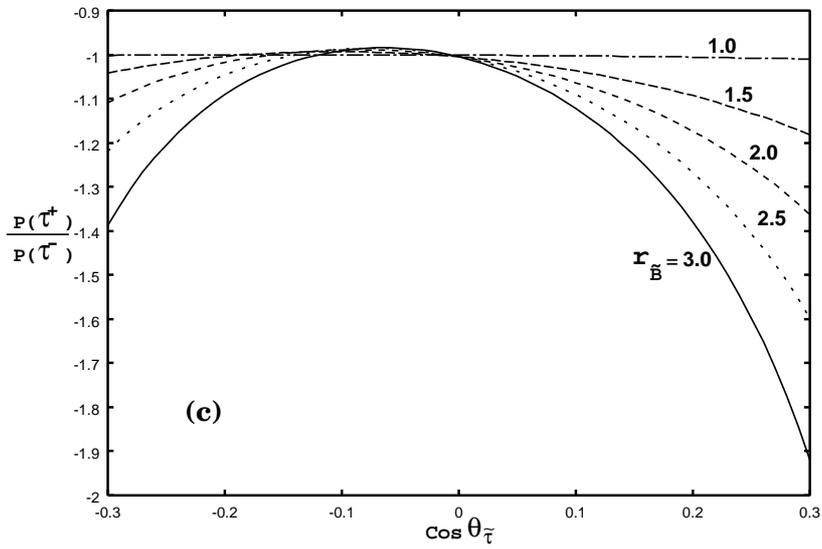
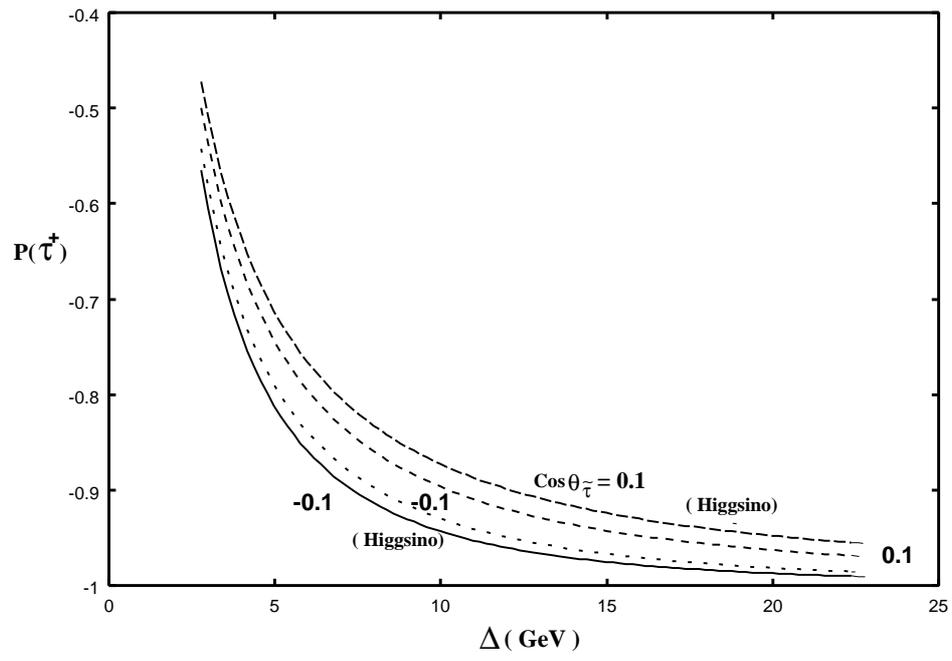
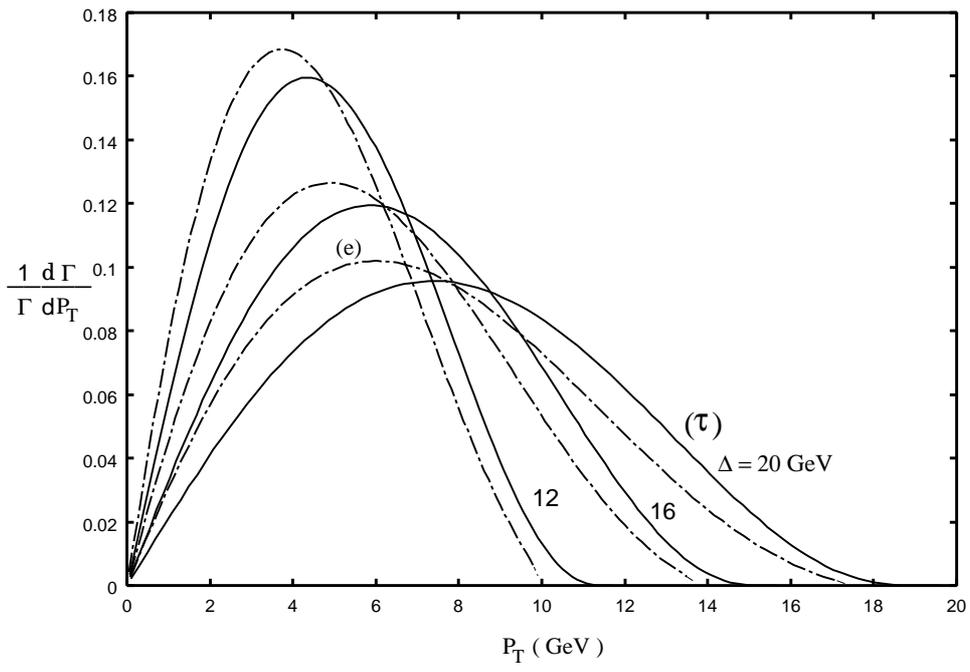
\end{document}